\documentstyle[12pt]{article}

\begin{document}

\begin{center}
{\large B.N.Kalinkin and F.A.Gareev\\
{\bf On the problem of synthesis of superheavy nuclei. \\
A short historical review on first theoretical predictions and new experimental
reality \\}}
\end{center}

\begin{sloppypar}
In connection with successful synthesis of a superheavy nucleus with
charge Z = 114 and mass number A =288, 289 performed in Dubna \cite{Og99}
it makes sense to recall theoretical studies in which for the first time
it has been predicted.

The problem is closely related to the experimental fact: nuclei with Z =
8, 20, 28, 50, 82 (for neutrons also N =126) are most stable to different
decay modes. This phenomenon can be interpreted  in the framework
of the shell model \cite{Pres} according to which the "magic" occupation numbers
are those of one-particle levels in nuclei after which a considerable
energy gap arises in the spectrum, and the binding energy gets maximal.
Consequently, a theoretical prediction of the existence of superheavy
nuclei beyond the periodic table should at least be based on
calculations of one-particle proton and neutron spectra aimed at
finding noticeable energy gaps in them.

To the mid sixties when this problem arose,  it became clear that the
widely used oscillator potential (the Nilsson scheme) is not valid for
that purpose. Perhaps, the only merit of it is that the wave functions
of one-particle states are rather simple. Physically, its essential
drawback is that the potential approaches to infinity near the surface of
a nucleus. As a result, the wave functions of one-particle spectrum
exhibit a wrong behaviour on the surface and periphery of a nucleus, i.e.
in the region that essentially contributes to the probabilities of
radiative transitions (the transition operator
$r^{\lambda}Y_{\lambda,\mu}(\theta,\phi)$, $\lambda=1,2,3,...$), elastic and
inelastic scattering and to those of other reactions. These computations
are by an order of magnitude larger or smaller than computations
based on "correct" wave
functions. A serious drawback of that scheme is a necessary change of
parameters of the potential and spin-orbital interaction when passing to
a higher shell. Therefore, no wonder that computations of the spectrum
of heavy nuclei based on the extrapolation of those parameters to
remote distances produced different magic values for Z and N. For example,
Nilsson et all \cite{Nils} obtained Z = 126 and N = 164, 184. Obviously
a scheme of that sort cannot be considered reliable, especially, for
predictions.

A reasonable solution of the problem may be based on a realistic
finite diffuse  potential $V_N(r)$ as a mean nuclear field and
on the justified form of the spin-orbital interaction \cite{Dav,Heiz}:
$$V_{SO}=\kappa(\vec{l}\bullet\vec{S})\frac{1}{r}\frac{dV_{N}}{dr},\eqno(1)$$
and also with
the charge distributed over a nucleus.  The most apt form of that potential
$$V_{N}(r)=-V_{0}(1+exp\frac{r-R_{0}}{a})^{-1},\;\;R_{0}=r_{0}A^{1/3},\eqno(2)$$
was proposed by Saxon and Woods \cite{Sax}.

In paper \cite{Kal} (Kalinkin B.N., Grabovskii Ya., Gareev F.A.  "On levels of
mean field of nuclei", JINR preprint  P--2682, 1966, the paper was submitted
to publication in Acta Physica Polonica on April 6, 1966 and accepted on
May 23, 1966)
we employed just this potential with parameters $V_0, r_0, a, k$
fixed from the data on low-lying levels of near-magic nuclei, on reactions of
one-nucleon transfers, elastic, inelastic scattering, and on polarization
effects \cite{Elt} --- \cite{Nem}. We developed an original method for numerical solution
of the Schroedinger equation with that potential and demonstrated its high
accuracy.

On June 16, 1966 we submitted for publication as a JINR preprint P--2793, 1966
and an article in the Phys.Lett. \cite{Gar}:
\begin{center}
A. Sobiczewski, F.A. Gareev, B.N. Kalinkin, "Closed shells for
$Z>82$ and $N>126$ in a diffuse potential well",
Phys. Lett. V.22, No 4(1966)500, received 22 July 1966, published 1 September
1966.
\end{center}

In this paper, based on the method elaborated in \cite{Kal}, we calculated
the proton and neutron energy levels versus A for $Z > 82$ and $N > 126$
for the Saxon--Woods potential with spin-orbital interaction. The
results show that possible magic numbers are Z = 114 and N = 184. The
computations were carried out with the parameters taken from ref.[8].
The solution turned out to be stable to variations of the parameters
of the potential and spin-orbital interaction caused by a possible
inaccuracy in their definition. No energy gap was observed in the system
of levels around Z = 126.

This paper has been the first publication in the available journals
giving a clear statement on possible existence of a superheavy nucleus
with Z = 114; it presents both the method of solution and demonstrates
the stability of the latter within the framework of a realistic potential
with justified values of parameters.

The importance of use of justified values of parameters for $V_{N}(r)$ and
$V_{SO}$ obtained from the analysis of the data on low-lying levels of
near-magic nuclei, on reactions of one-nucleon transfers, elastic, inelastic
scattering, and on polarization effects \cite{Elt}-\cite{Nem} demonstrated by the results
of \cite{Wong}. The parameter $a$  in the potential $V_{N}$ \cite{Wong} has been used
unjustified large magnitude (see for details \cite{Wong}) which changed strongly
spectrum of nuclei and lead to the magic number $Z=126$

The JINR preprint P -- 2793 was then distributed by N.I.Pyatov among
participants of the Int.Symposium on Why and how should we investigate
NUCLIDES FAR OFF THE STABILITY LINE, Lysekil, Sweden, August 21--27, 1966,
where the considered problem was of common interest (Session IX:
Nucleosynthesis; Chairman: W.J.Swiatecki).

In the report \cite{Meld67} by
H. Meldner: "Predictions of new magic regions and masses for super-heavy
nuclei from calculations with realistic shell model single particle
hamiltonians",
Proc. of the Intern.  Lyseki Symposium,  Sweden, August 21-27, 1966.
Received 14 September 1966, published 18 October 1967, Ark. Fys. 36(1967)593.

H.Meldner informed that new magic numbers should be Z = 114 and N = 184
and at the end made the comment: "Note added in proof. In the meantime the
proton shell Z = 114 has been found in independent investigations [13]".
Reference [13] of that report  is \\
{\sl [13] Nilsson S.G., private communications, Strutinsky V.M., private
communications, Sobiczewski A., Gareev F.A., Kalinkin B.N. (preprint).}

So, when H.Meldner submitted his report on September 14, 1966, he already
had our preprint.

We consider also that our studies and studies by H.Meldner were carried
out independently. However, we do not agree with G.Herrmann, the author
of recent paper \cite{Her}, from which it may be concluded that  it was just
H. Meldner who first predicted magic numbers Z = 114 and N = 184. Let
us discuss this question in greater detail.

In \cite{Meld67}, p. 595, H. Meldner reported:

{\sl The same result was obtained in simpler calculations with local potentials
two years ago (9).}\\
{\sl (9) see discussion on super-heavy nuclei in W.D. Myers and W.J. Swiatecki,
Nucl. Phys. 81,1 (1966); or UCRL-11980(1965), based on calculations quoted
there under ref. (23)}

In paper by W.D. Myers and W.J. Swiatecki,
Nucl. Phys. 81,1 (1966), ref. (23) looks as follows:\\
{\sl (23) H. Meldner and P. Roper (Institut fur Theoretische Physik der
Universitat Frankfurt/M.), personal communication (1965)}

We quote a brief fragment from that paper (pp. 49, 50 ): "In our mass
formula we have included, for purposes of illustration, magic numbers at
$Z=126$ and $N=184, 258$ --- see fig 19.( the latter numbers are obtained by
following the sequence of major shells in harmonic oscillator potential with
spin-orbit coupling). We do not wish to imply that there are grounds for
believing that any of these magic numbers would show up in practice, and we
use them only to illustrate that some of the consequences would be if a magic
number turned out to be present in the general neighbourhood of super-heavy
nuclei somewhat beyond the end of the periodic table.The actual values of the
magic numbers might be different; for example, we have recently learned (23)
that $Z=114,\;\;N=184$ is a possible candidate for a doubly magic nucleus...
What we wish to point out is that if a (double) magic number exists, then an
important consideration affecting the possible stability of the corresponding
nucleus is the considerable increase in the barrier against fission and,
consequently, in the spontaneous fission half-life.

... In order to proceed in a realistic manner with discussion of the existence
and location of possible islands of stability beyond the periodic table the
first requirement is the availability of estimates for the location and strength
of magic number effects in that region. When such estimates have become
available (through single-particle calculations in realistic nuclear potentials)
it will be possible to apply our semi-empirical treatment of nuclear masses
and deformations to the predictions of the fission barriers of hypothetical
super-heavy nuclei..."

>From the above quotations it follows that

First, W.D. Myers and W.J. Swiatecki in their calculations used the values
of magic numbers obtained by other authors with the use of harmonic
potential. Estimates on the basis of realistic potentials
were not available for them at that time.

Second, they obtained information on a possible realization of the double
magic nucleus with $Z=114,\;\;N=184$ from a personal communication of H.
Meldner and P.Roper who did not published them anywhere, which is verified
by the absence of any reference to that work in the report \cite{Meld67}.

It is obvious that personal communications cannot be reason of the priority.
The priority requires official publications of results with the method they
have been obtained, accuracy, and stability of the solution permitting
verification of the results by any physicist.

It remains to declare that the report by H. Meldner \cite{Meld67} is his first
official communication on possible existence of the nucleus with $Z=114$.

The following two reports are also devoted to realization of superheavy
nuclei.

In the report \cite{Gust}:\\
C. Gustafson, I.L. Lamm, B. Nilsson, S.G. Nilsson, "Nuclear deformations
in the rare-earth and actinide regions with excursions off the stability line
and into the super-heavy elements",
Received 14 September 1966, published  in Ark. Fys. 36(1967)613.\\
it is stated that
{\sl as a by-product of these computations it appears reasonable to forecast
that the "magic" proton candidate is $Z=114$ and not $Z=126$ while for neutrons
$N=184$ is a rather questionable "magic" number. These predictions remain valid
also when a reasonable extrapolation is made in the values $\mu$ and $\xi$
(Fig. 5, cf. ref. (8))\\
(8) Sobiczewski, Gareev and Kalinkin, to appear in Nucl. Phys.}

In the report \cite{Str66}:\\
V.M. Strutinsky, "Microscopic calculations of the nucleon shell effects
in the deformation energy of nuclei",\\
Received 14 September 1966, published  in Ark. Fys. 36(1967)629.\\
The behavior of deformation energy is studied for some heavy and
superheavy nuclei with consideration for shell effects. Use is made
for the "Nilsson scheme" (a traditional version). The most stable nucleus
has been that with $126^{310}$. Possible realization of a nucleus with
$Z=114$ is not discussed.

Next "burst" of the activity in discussing the existence of superheavy
nuclei took place at the International conference on the physics of heavy
nuclei held at Dubna on October  13--19, 1966. There two reports were
delivered \cite{Str67,Fri67}:\\V.M.Strutinskii and Yu.A.Muzychka
"Some shell effects in transuranium nuclei",\\A.M. Friedman,
"Calculations on the production of the next closed shell nucleus and other
nuclei".

Based on a realistic potential, the authors conclude that
$Z=114$ and $N=184$ are the most pronounced magic numbers in the
region of superheavy nuclei. Also, both the reports do not refer to our
work \cite{Gar}. Proceedings of that conference were published on October
16, 1967.

Concluding a brief review of studies made in 1966 and devoted to the
possibility of existence of a heavy nucleus with $Z=114$, we note once
more that it has first been predicted in our work \cite{Gar}.

It is also important to recall that our method of solving the problem
\cite{Gar} was later verified by V.A.Chepurnov \cite{Chep} who
reproduced our results by direct numerical solution with a high accuracy.
Also, we generalized it to a realistic nuclear field for strongly deformed
nuclei \cite{Gar68}-\cite{Gar71}. Practical application of the
generalized method in a lot of investigations on the spectroscopy of the
rare-earth and transuranium group carried out at the BLTP, JINR in recent
years (see, e.g., monographs \cite{Sol71,Sol81})
has proved its high efficiency. Therefore, we may hope, it could be used
for studying superheavy deformed nuclei of the island of stability whose
actual synthesis begins just now.

Evidence for the island of stability to exist rather than a single
superheavy nucleus, to our mind, comes from the logic of \cite{Gar}
the very fact of synthesis of
the nucleus with the magic number of protons $Z=114$ and nonmagic number
of neutrons $N=175$.  If so, then stable should be both the doubly magic
nucleus with $Z=114$ and $N=184$ (the island center) and the nucleus with
the nonmagic number of protons $Z>114$ and magic number of neutrons
$N=184$.  Nuclei with $Z$ and
$N$ near the above-mentioned combinations should also be stable \cite{Kal99}.

So, the theoretical prediction of a superheavy nucleus with $Z=114$,
formulated for the first time at Dubna \cite{Gar}, that has allowed
a goal-directed experimental search has been testified by its actual
synthesis also at Dubna many years later.

In conclusion, we note that  the synthesis of
superheavy nucleus $^{289}114$ was further developed:
new heaviest nuclides $^{288}114$ and $^{284}112$ were observed
\cite{Og99a}. An attempt of synthesis of the superheavy nuclei was performed
in Berkeley \cite{Nin99}. This attempt was commented in \cite{Og99a}:
"{\sl The synthesis of $^{293}118$ and its sequential $\alpha$-particle emission to
the daughter isotopes with $Z=116-106$ in the bombardment of $^{208}Pb$ with
$2.3*10^{18}$ 449-MeV $^{86}Kr$ ions using the Berkeley separator BGS was
announced in April-May, 1999. Three decay chains were observed, each consisting
of an implanted  atom and six subsequent $\alpha$-decays. Another experiment
with this reaction was carried out at the same bombarding energy at GSI, in
Darmstadt, using the separator SHIP. No correlated $\alpha$-decay chains were
observed yet, with a similar beam dose of $2.9*10^{18}$ Kr ions \cite{Hof99}.}"
Therefore, the results of analyzing this reaction is to be continued.

Thus, the experimental research of the island of stability for superheavy
nuclei was started with a high activity.

\end{sloppypar}


\begin{thebibliography}{99}

\bibitem{Og99} Yu. Ts. Oganessian et al., Preprint JINR E-7-99-53, Dubna, 1999;
JAF, {\bf V.63}, 1769, 2000; Phys. Rev., C62.0411604(R), 2000; Phys. Rev. Lett.,
{\bf 83}, 3154, 1999; Nature (London) {\bf 400}, 242, 1999; Eur. Phys. J. A5,
63, 1999.
\bibitem{Pres} M.A. Preston, Physics of Nucleons. Addison-Weslag Pub. C.,
Jnc., Reading, Massachusetts, Palo-Alto-London, 1962.
\bibitem{Nils} S.G. Nilsson, Kgl Danske Vidensk, Selsk. mat-fys. Medd. {\bf 22},
N 16 (1955); B.R. Mottelson, S.G. Nilsson, Kgl Danske Vidensk,
Selsk. mat-fys. kr. {\bf 1}, N 8 (1959).
\bibitem{Gar} F.A. Gareev, B.N. Kalinkin, A. Sobiczewski, JINR P-2793, 1966;
Phys. Lett. {\bf 22}(1966)500.
\bibitem{Dav} A.S. Davydov, Teoriya atomnogo yadra. Fiz.-mat. GIZ, Moskow, 1958.
\bibitem{Heiz} W. Heizenberg, Teorie des Atomkerns, Gottingen, 1951.
\bibitem{Sax} D. Saxon, R. Woods, Phys. Rev. {\bf 95}(1954)577.
\bibitem{Kal} B.N. Kalinkin, Ya. Grabovskii, F.A. Gareev, JINR P-2682, 1966;
Acta Physica Polonica, {\bf XXX}, N6(1966)999.
\bibitem{Elt} L.R.B. Elton, Nuclear sizes. Oxford University Press, 1961.
\bibitem{Hods} P.E. Hodgson, The optical model of elastic scattering. Oxford,
The Clarendon Press, 1963.
\bibitem{Lev} I.I. Levintov, Physica {\bf XXII}(1956)1178; JETP, {\bf 30}
(1958)987.
\bibitem{Nem} P.E. Nemirovskii, Sovremennye modeli atomnogo yadra. Atomizdat,
Moscow, 1960.
\bibitem{Wong} C.Y. Wong, Phys. Lett. {\bf 21}(1966)688.
\bibitem{Meld67}. H. Meldner,
Proc. of the Intern.  Lysekil Symposium,  Sweden, August 21-27, 1966.
Received 14 September 1966, published 18 October 1967, Ark. Fys. {\bf 36}
(1967)593.
\bibitem{Her} G. Herrmann, Nuclear Physics News {\bf 8}, No 2 (1998)7.
\bibitem{Gust} C. Gustafson, I.L. Lamm, B. Nilsson, S.G. Nilsson,
Received 14 September 1966, published 18 October 1967, Ark. Fys. {\bf 36}
(1967)613.
\bibitem{Str66} V.M. Strutinsky,
Received 14 September 1966, published 18 October 1967, Ark. Fys. {\bf 36}
(1967)629.
\bibitem{Str67} V.M. Strutinsky, Yu.A. Muzychka,
in Proceedins of International Conference on Heavy Ion Physics, Dubna,
13-19 October 1966., recieved on 16 October
1966., published in November 1967., p. 51.
\bibitem{Fri67} A.M. Friedman,
in Proceedins of International Conference on Heavy Ion Physics, Dubna,
13-19 October 1966., recieved on 16 October
1966., published in November 1967., p. 39.
\bibitem{Chep} V.A. Chepurnov, Soviet Journal of Nuclear Physics {\bf 6}
(1967)955.
\bibitem{Gar68} F.A. Gareev, S.P. Ivanova, B.N. Kalinkin, Acta Physica Polonica
{\bf 32} (1967) 461; {\bf 33} (1968) 133; Izv. AN SSSR, ser. fiz., {\bf 32}
(1968) 1690.
\bibitem{Gra} Ya. Grabovskii, B. Kalinkin, Nucleonica {\bf XIV} (1967) 571.
\bibitem{Gar76} J. Bang, F.A. Gareev, I.V. Pusynin, R.M. Jamalejev, Nuclear
Physics {\bf A261}(1976)59.
\bibitem{Gar78} …. Bang, F.A. Gareev, S.P. Ivanova, Particles and Nuclei
{\bf 9}(1978)286.
\bibitem{Gar69} F.A. Gareev, M.I. Chernei, S.P. Ivanova,
Soviet Journal of Nuclear Physics {\bf 9}(1969)308.
\bibitem{Gar68a} F.A. Gareev, B.N. Kalinkin, N.I. Pyatov, M.I. Chernei,
Soviet Journal of Nuclear Physics {\bf 8}(1968)305.
\bibitem{Gar72} H. Shulz, H.J. Wiebicke, F.A. Gareev, Nucl. Phys. {\bf A180}
(1972)625.
\bibitem{Gar71} F.A. Gareev, S.P. Ivanova, L.A. Malov, V.G. Soloviev,
Nucl. Phys. {\bf A171}(1971)3.
\bibitem{Sol71} V.G. Soloviev, Teoriya sloznyh yader.  Nauka, Moscow, 1971.
\bibitem{Sol81} V.G. Soloviev, Teoriya sloznyh yader. Yadernye modeli.
Energoizdat, Moscow, 1981.
\bibitem{Kal99} B.N. Kalinkin, F.A. Gareev, Preprint JINR E7-99-107, Dubna, 1999.
\bibitem{Nin99} V. Ninov et al., Phys. Rev. Lett. {\bf 83},1104(1999).
\bibitem{Og99a} Yu. Ts. Oganessian et al., Preprint JINR E-7-99-347, Dubna, 1999;
\bibitem{Hof99} S. Hofman et al., private communication.
\end{thebibliography}
\end{document}